\documentstyle[psfig]{mn}

\newif\ifAMStwofonts
\AMStwofontstrue

\ifoldfss
  \ifCUPmtlplainloaded \else
    \NewTextAlphabet{textbfit} {cmbxti10} {}
    \NewTextAlphabet{textbfss} {cmssbx10} {}
    \NewMathAlphabet{mathbfit} {cmbxti10} {} 
    \NewMathAlphabet{mathbfss} {cmssbx10} {} 
  \fi
  \ifAMStwofonts
    \ifCUPmtlplainloaded \else
      \NewSymbolFont{upmath} {eurm10}
      \NewSymbolFont{AMSa} {msam10}
      \NewMathSymbol{\upi}     {0}{upmath}{19}
      \NewMathSymbol{\umu}     {0}{upmath}{16}
      \NewMathSymbol{\upartial}{0}{upmath}{40}
      \NewMathSymbol{\leqslant}{3}{AMSa}{36}
      \NewMathSymbol{\geqslant}{3}{AMSa}{3E}

      \let\leq=\leqslant 
       
    \fi
  \fi
\fi 

\ifnfssone
  \newmathalphabet{\mathit}
  \addtoversion{normal}{\mathit}{cmr}{m}{it}
  \addtoversion{bold}{\mathit}{cmr}{bx}{it}
  \newmathalphabet{\mathbfit} 
  \addtoversion{normal}{\mathbfit}{cmr}{bx}{it}
  \addtoversion{bold}{\mathbfit}{cmr}{bx}{it}
  \newmathalphabet{\mathbfss} 
  \addtoversion{normal}{\mathbfss}{cmss}{bx}{n}
  \addtoversion{bold}{\mathbfss}{cmss}{bx}{n}
  \ifAMStwofonts
    \ifCUPmtlplainloaded \else
      \UseAMStwoboldmath
      \makeatletter
      \new@mathgroup\upmath@group
      \define@mathgroup\mv@normal\upmath@group{eur}{m}{n}
      \define@mathgroup\mv@bold\upmath@group{eur}{b}{n}
      \edef\UPM{\hexnumber\upmath@group}
      \new@mathgroup\amsa@group
      \define@mathgroup\mv@normal\amsa@group{msa}{m}{n}
      \define@mathgroup\mv@bold\amsa@group{msa}{m}{n}
      \edef\AMSa{\hexnumber\amsa@group}
      \makeatother
      \mathchardef\upi="0\UPM19
      \mathchardef\umu="0\UPM16
      \mathchardef\upartial="0\UPM40
      \mathchardef\leqslant="3\AMSa36
      \mathchardef\geqslant="3\AMSa3E

      \let\leq=\leqslant 

    \fi
  \fi
\fi 

\ifnfsstwo
  \DeclareMathAlphabet{\mathbfit}{OT1}{cmr}{bx}{it}
  \SetMathAlphabet\mathbfit{bold}{OT1}{cmr}{bx}{it}
  \DeclareMathAlphabet{\mathbfss}{OT1}{cmss}{bx}{n}
  \SetMathAlphabet\mathbfss{bold}{OT1}{cmss}{bx}{n}
  \ifAMStwofonts
    \ifCUPmtlplainloaded \else
      \DeclareSymbolFont{UPM}{U}{eur}{m}{n}
      \SetSymbolFont{UPM}{bold}{U}{eur}{b}{n}
      \DeclareSymbolFont{AMSa}{U}{msa}{m}{n}
      \DeclareMathSymbol{\upi}{0}{UPM}{"19}
      \DeclareMathSymbol{\umu}{0}{UPM}{"16}
      \DeclareMathSymbol{\upartial}{0}{UPM}{"40}
      \DeclareMathSymbol{\leqslant}{3}{AMSa}{"36}
      \DeclareMathSymbol{\geqslant}{3}{AMSa}{"3E}

      \let\leq=\leqslant 

    \fi
  \fi
\fi 

\ifCUPmtlplainloaded \else
  \ifAMStwofonts \else 
    \def\upi{\pi}
    \def\umu{\mu}
    \def\upartial{\partial}
  \fi
\fi

\title{Confusion limit due to galaxies: using {\em SIRTF}'s 
       Infrared Array Camera}

\author[P. V\"ais\"anen, E.V. Tollestrup \& G.G. Fazio]
       {Petri V\"ais\"anen,$^{1,2,}$\thanks{E-mail: vaisanen@astro.helsinki.fi} 
       Eric~V.~Tollestrup,$^2$ and Giovanni~G.~Fazio$^2$ \\
$^1$Observatory, P.O.B. 14, 00014 University of Helsinki, Finland \\
$^2$Harvard-Smithsonian Center for Astrophysics, 60 Garden St., Cambridge, MA 02138, USA}

\date{Accepted April 2001}

\pagerange{\pageref{firstpage}--\pageref{lastpage}}
\pubyear{2000}

\begin{document}

\maketitle

\date{\today}

\label{firstpage}

\newcommand{\ea}{{et~al.~}}
\newcommand{\eg}{{e.g.~}}
\newcommand{\ie}{{i.e.~}}
\newcommand{\apj}{{ApJ}}
\newcommand{\mnras}{{MNRAS}}

\begin{abstract}

Recent {\em ISO}-data has allowed for the first time observationally 
based estimates for source confusion in mid-infrared surveys.
We use the extragalactic source counts from ISOCAM in conjunction 
with $K$-band counts to predict the confusion due to galaxies in deep 
mid-IR observations.   
We specifically concentrate on the near-future Space Infrared Telescope
Facility ({\em SIRTF}) mission, and 
calculate expected confusion for the Infrared Array Camera (IRAC) 
onboard {\em SIRTF}.  A defining scientific goal of the IRAC 
instrument will be the study of high redshift galaxies using a deep, confusion 
limited wide field survey at 3-10 $\umu$m.  A deep survey can reach 
3 $\umu$Jy sources with reasonable confidence in the shorter wavelength 
IRAC bands.  Truly confusion limited images with the $8\umu$m will be 
difficult to obtain due to practical time constraints, unless infrared 
galaxies exhibit very strong evolution beyond the deepest current observations.
We find $L^\star$ galaxies to be detectable to $z$=3--3.5 at 8 $\umu$m, which
is slightly more pessimistic than found by Simpson \& Eisenhardt (1999).
\end{abstract}

\begin{keywords}
infrared: galaxies -- cosmology: observations -- galaxies: evolution -- 
galaxies: photometry -- methods: observational -- techniques: photometric
\end{keywords}

\section{Introduction}

One of the successes of the Infrared Space Observatory ({\em ISO}) mission 
have been the new extragalactic 
source counts in the mid-infrared, extracted from both deep 
(Oliver et al. 1997, Elbaz et al. 1999) 
and wide field (Serjeant et al. 2000) surveys.
These counts are important in planning effective observing strategies for
the next generation of space-borne IR missions, the {\em SIRTF} 
(Fanson et al. 1998)
and {\em ASTRO-F} (Pearson et al. 2000), particularly with respect to expected
source densities and confusion. 

Confusion is usually defined as the fluctuations of the background sky 
brightness below which sources cannot be detected individually -- these 
fluctuations are caused by intrinsically discrete extragalactic sources 
(we do not consider confusion due to Galactic cirrus in this work). 
Thus the fluctuations, or noise, come from the same sources one wishes to 
study.  

Confusion sets an important and fundamental limit to astronomical observations.
For a given wavelength and resolution, it will not be 
productive to extend the exposure time indefinitely since a flux density 
limit will be reached beyond which no additional discrete sources can be 
extracted.  This is the limit usually referred to as the {\em confusion limit}.
Confusion cannot be avoided even with arbitrarily high resolution 
because of the finite sizes of sources on the sky.

Just as new sub-mm data allowed Blain, Ivison \& Smail (1998) to set 
observational limits on source confusion at those wavelengths,
we are now able to estimate more accurately the
confusion at mid-infrared wavelengths using deep galaxy counts from {\em ISO} 
data.  Confusion in mid-IR has been discussed in connection with the 
individual deep surveys (\eg Oliver \ea 1997 for the {\em ISO} HDF survey).

A useful theoretical 
definition of confusion was developed by Scheuer (1957) and 
followed up in numerous other works.  More recently, essentially the same 
formalism has been used to predict confusion in observations done at radio
(\eg Franceschini 1989, Wall \ea 1982, Condon 1974), sub-mm (Toffolatti \ea
1998), and X-ray (Barcons 1992, Scheuer 1974) wavelengths.  It turns out that 
sources at the flux density level where the surface density is 
approximately 1 source per beam produce the bulk of the confusion noise.  
This is easy to understand: at higher surface densities the multitude of 
sources per beam suppresses the fluctuations (contributing more to the 
absolute background level), while at lower surface density levels the many 
'empty' beams also decrease the fluctuations.

However, confusion is not a very straightforward limit since it is a function 
of wavelength, resolution, and other instrument response dependent factors.  
Moreover, calculations of confusion limits for a given instrument are 
highly source count -model dependent.  `Confusion' is also sometimes 
confused with source detection probabilities, \ie completeness level 
estimation when performing source extraction and photometry.  Bright 
sources are often neglected from confusion calculations, but these 
bright objects can be a major factor in lowering the completeness of 
detections in a given survey.  The aspect of completeness vs.\ confusion 
is perhaps 
the least-studied in the otherwise extensive literature on confusion.

The goal of this paper is two-fold: i) to study the relation of confusion 
noise to actual detection of sources, and thus arrive at a practical 
definition for confusion, and ii) to predict this confusion limit in mid-IR 
specifically for the forth-coming space infrared mission, the {\em SIRTF}.  

We start from analytical and theoretical confusion determination and proceed 
to a more observer oriented completeness analysis.  
In particular, we will define confusion `noise' and confusion `limit' 
and investigate their relation to completeness levels in source extraction.  
We specify when the different methods of deriving confusion are most 
appropriate and present both methods for IRAC.  
The relevant quantitative numbers 
for confusion with IRAC/{\em SIRTF} are presented in 
Section~\ref{disc} and Table 1.

\section{Analytical solution for confusion noise}
\label{analytical}

Given the source-count distribution and the response pattern of the detector, 
we can estimate the confusion analytically -- this has been done by many 
authors (see above), here we mainly follow the notation of Franceschini \ea 
(1989, 1991).

Let $N(S)$ be the differential distribution of sources of flux $S$ ($N(S)$ is 
used for differential counts throughout this paper instead of $dN/dS$).  
We assume that the sources are distributed randomly on the sky. 
Now let us include  the information of the observing instrument:  
let $f(\theta, \varphi)$ be the point-spread profile (normalized to unity) of 
the detector.  Then the amplitude of the response at location 
$(\theta , \varphi)$ from the beam axis, is 
\begin{equation}
x=S\cdot f(\theta ,\varphi).
\end{equation}
The mean number of responses, $R(x)$,  
with amplitudes between $x$ and $x+dx$ in a solid angle $d\Omega$ is:
\begin{equation}
R(x) dx = \int_{\Omega_{b}} N\left( \frac{x}{f(\theta, \varphi)} \right) 
\frac{d\Omega}{f(\theta ,\varphi)} \, dx ,
\label{eq2}
\end{equation}
where we integrate over the whole beam $\Omega_{b}$. 

The intrinsic fluctuations of $R(x)$ can be written simply as the second 
moment $\sigma_{\rm conf}$ of the $R(x)$ distribution:
\begin{equation}
\sigma_{\rm conf}^{2} = \int_{0}^{x_{c}} x^{2} R(x) dx.
\label{eq3}
\end{equation}

The lower integration limit is 0 because $S$ is always positive.  The upper 
limit is not so straightforward; the variance would diverge if high-flux 
truncations are not introduced in either the source counts $N(S)$ or the 
response distribution $R(x)$.  The limit is often set to $x_{c} = Q\sigma$, 
with $Q$ in the range of 3--5.  This kind of upper limit makes the 
calculation an iterative process.  The logic of having a {\em response} 
cut-off rather than a cut-off at specific flux $S$, is that the latter 
would eliminate the contribution by bright sources alltogether whereas 
in reality their presence -- even relatively far away from the beam 
center -- could significantly affect the measurement due to the wings 
of their PSFs.  

Notice that if $Q=3$, for example, 
it does not follow that the objects detected 
will be `3$\sigma$'-detections in the conventional sense. The $Q$-value 
merely tells how large responses are included in the calculation of the 
confusion: this corresponds to setting a limit below which a response 
would not be detected with sufficient confidence as such.  Thus for 
Poissonian and Gaussian distributions $Q=3$ corresponds to a probability 
$\leq 0.3$ per cent 
that a response $x$ results from a large fluctuation of the 
background.  The $\sigma_{\rm conf}$ value from Eq.~\ref{eq3} must be 
multiplied by 3 to get the `3$\sigma$ confusion limit' for the detection 
of objects regardless of the $Q$ used.

For simplicity we will assume a Gaussian beam pattern 
\begin{equation}
f(\Psi) = \exp(4 \Psi \ln 2).
\label{eq1}
\end{equation}
where $\Psi = (\theta/\theta_{0})^{2}$.
Performing the integration over the solid angle $\Omega_{b}$,  
Eq.~\ref{eq3} can be written as 
\begin{equation}
\sigma^{2}_{\rm conf}(x_{c}) = \pi \theta_{0}^{2} I(x_{c}), 
\label{eq5}
\end{equation}
where 
\begin{equation}
I(x_{c}) = \int_{0}^{x_{c}} x^2 dx \int_{0}^{\infty} 
N\left(\frac{x}{f(\Psi)}\right)  \exp (4\Psi \ln 2) d\Psi.
\label{eq6}
\end{equation}
Because of the integration limit ($x_{c} = Q\sigma$) it is 
easier to compute $\theta_{0}$ (the FWHM of the beam) as a function of 
$\sigma$, \ie 
\begin{equation}
\theta_{0} = \sigma / (\pi I(x_{c}))^{1/2}
\label{eq6-5}
\end{equation}
and then determine the confusion for a given $\theta_{0}$.

If the source counts have a power-law form, $N(S) = k S^{\gamma}$,
Eq.~\ref{eq5} can be integrated to give 
\begin{equation}
\sigma_{\rm conf} = ( \frac{k \Omega_e}{3-\gamma} )^{1/2} x_c^{(3-\gamma) / 2}
\label{eq7}
\end{equation}
where $\Omega_e = \int f(\Psi) ^{\gamma -1} d\Omega$ is the effective beam size
(\eg Condon 1974, Hacking \& Houck 1987).  If the Gaussian
beam is assumed, as above, we can find a relation
\begin{equation}
\sigma_{\rm conf} \propto \theta_{0}^{2/(\gamma-1)}
\label{eq7-5}
\end{equation}
in the absence of cut-offs $S_{c}$ in source fluxes.

In summary, 
it is evident that confusion will be a function of 
the beam size, the source counts, cut-offs related 
to the source and response distribution, and the $Q$-term:
\begin{equation}
\sigma_{\rm conf} = \sigma_{\rm conf}(\theta_{0},N(S),x_{c},S_{c},Q)
\end{equation}

\section{The IRAC instrument}

The Infrared Array Camera, (Fazio et al.\ 1998), 
is one of the three focal plane instruments
on board {\em SIRTF}.  IRAC is a four-channel camera that provides 
simultaneous images at 3.6, 4.5, 5.8, and 8 $\umu$m.  Two adjacent
5.12\arcmin $\times$ 5.12\arcmin\ fields of view are seen by these four
channels in pairs.  All four detectors are $256\times256$ pixels, with
1.2\arcsec/pixel scale.  The two short wavelength channels use InSb 
detectors while the two longer wavelengths have Si:As IBC detectors.

In the following, we concentrate on the shortest and longest wavebands, the
3.6 and 8 $\umu$m channels.  The sensitivities in these two bands and 
characteristics of
galaxy models, as well as confusion estimates bracket
those expected from the two intermediate channels.  
The expected $5\sigma$ sensitivities in 200 s for the
3.6 and 8 $\umu$m bands are 3.1 and 23.2 $\umu$Jy, respectively,
for point sources, and 0.9 and 5.1 $\umu$Jy/arcsec$^2$ for extended sources.
The bandwidths are 21 and 38 per cent, respectively.

The major scientific objectives for IRAC and {\em SIRTF} are i) to study 
the early universe; ii) to study ultraluminous galaxies
and active galactic nuclei; iii) to search for and study brown dwarfs and
superplanets, and iv) to discover and study protoplanetary and planetary
debris disks.  These objectives are also mirrored in the six large `SIRTF
Legacy Science Programs', which will utilize nearly 3200 hours of 
{\em SIRTF}
observing time primarily during the first year of the mission.  The projects
are: `The {\em SIRTF} Galactic Plane Survey'; ´GOODS: Great Observatories 
Origins Deep Survey'; `From Molecular Clouds to Planets'; `SINGS: The 
{\em SIRTF} Nearby Galaxies Survey -- Physics of the Star-Forming ISM
and Galaxy Evolution'; `SWIRE: The {\em SIRTF} 
Wide-area Infrared Extragalactic 
Survey'; and `The Formation and Evolution of Planetary Systems: 
Placing Our Solar System in Context'.  All of these are expected to result
in substantial databases that will be invaluable for archival research, 
and in planning subsequent programs on {\em SIRTF} and on other space-borne, 
airborne, and ground-based observatories.

While the motivation of this paper is connected
to deep surveys of the high redshift universe (\eg GOODS), confusion limit
due to galaxies is a factor also in the deeper brown dwarf searches.
Additionally, confusion due to stars becomes important in these 
surveys when operating close to the Galactic plane, at $b<20$\degr
(\eg the Galactic Plane Survey project).

\section{Source count models}
\label{models}

\subsection{Baseline model}

We explore the effects of confusion and completeness 
with simple models of mid-infrared galaxy counts.  
As a first step, observed near-infrared $K$-band source counts are 
extrapolated to mid-IR.  

To predict the counts at 
another wavelength, the specifics of the observing filter are needed, as well
as the shape of the spectral energy distribution (SED) of the contributing
sources at different redshifts and the numbers of these sources. The idea 
is to construct a model which fits reasonably well the observed counts, 
and then 
calculate the counts of the target band with the same model. 
Thus, a consistent extrapolation requires knowledge of the 
SEDs of different galaxy types, the local galaxy 
luminosity functions (LF) of the types, possible evolution of the sources, 
and a cosmological model.  The observed counts
themselves can then be shifted to the other band using information of the 
predicted colors as a function of magnitude.

In the following the {\em baseline model} is
an extrapolation, such as described above, constructed to  
fit observed $K$-band counts. 
To compute the galaxy counts we used the general formalism presented earlier in
V\"ais\"anen (1996).  For the SEDs and luminosity evolution of galaxies 
we use the `bc96' models (Bruzual \& Charlot 1993) 
as presented in Gardner (1998). 
We also made use of {\em ncmod}, a useful general purpose galaxy number 
count model presented in Gardner (1998).   
We adopt a prescription for internal absorption by dust in 
galaxies from therein as well as the included six different galaxy types 
(passively
evolving E/S0, Sab, Sbc, Scd, Irr, and a small population of constantly 
star-forming dwarfs).  
The galaxy mix is from Yoshii \& Takahara (1988) and
the $K$-band local luminosity function (same for
all galaxy types) from Gardner \ea (1997).  
For cosmology we used $H_{0} = 50 {\rm km} 
{\rm s}^{-1} {\rm Mpc}^{-1}$ and $q_{0} = 0.02$.

\begin{figure}
\centerline{\psfig{figure=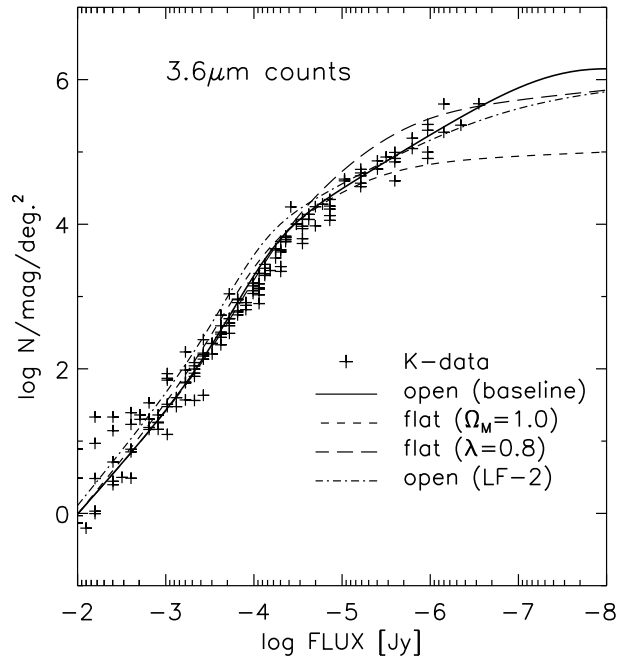,width=9.5cm} }
\caption{Simulated IRAC 3.6 $\umu$m band counts. 
The crosses show actual $K$-band galaxy counts 
shifted to the shortest wavelength IRAC band.  The model used in
the transformation includes pure 
luminosity evolution, uses Gardner et al.\ (1996) 
$K$-band local luminosity function, and has open cosmology. 
The calculated counts of the
same model are plotted as the thick solid line.  The lowest dashed line
shows counts from the same model but with flat $q_{0}=0.5$ cosmology, 
whereas the
long-dashed curve uses flat, cosmological constant dominated cosmology 
($\lambda_{0}=0.8$).  The dash-dot line is calculated using an alternative LF,
that of Szokoly et al.\ (1998), which results in a slightly 
higher normalization at the bright end.}
\label{model3_counts}
\end{figure}

Using the model above, we shift the observed $K$-band counts,
a compilation from V\"ais\"anen et al.\ 2000, 
to 3.6 $\umu$m -- these are shown as crosses in Fig.~\ref{model3_counts}. 
The predicted source count of the same model is overplotted as the thick solid
line.  This same baseline model is 
calculated also for 6.7 and 8$\umu$m bands.
The baseline model corresponds to the one in Gardner's (1998) Figure 11. 
There are other models overplotted in Fig.~\ref{model3_counts},
corresponding to different cosmologies.  The dash-dot curve 
uses the local near-IR luminosity function of galaxies from 
Szokoly \ea (1998), which we have found to predict near-IR counts 
fitting well our own wide-field galaxy counts (V\"ais\"anen \ea 2000).

\subsection{{\em ISO}-data based model}

For the 5.8 and 8 $\umu$m predictions, 
the mid-IR spectral properties of galaxies become important and 
extrapolations such as used for the baseline model
are not trustworthy.  This is because the `bc96' SEDs 
used above do not include the broad emission features in the 
spectra of galaxies between 3 and 12 $\umu$m,
commonly thought to be the signature of Polycyclic Aromatic Hydrocarbons (PAH).
While global properties of galaxy populations
at mid-IR remain yet uncertain, recent {\em ISO} data certainly has shown 
that PAH features are an
important factor which cannot be ignored (eg.\ Mattila, Lehtinen \& 
Lemke 1999, Genzel \& Cesarsky 2000).
This is particularly true for any dusty and infrared bright objects, 
which might well dominate counts at mid-IR.  
Modelling of the relevant dust and re-radiation processes is complicated
(see eg.\ Guiderdoni \ea 1998, Silva \ea 1998, Xu \ea 1998); therefore, 
to be as observationally based as possible,
our chosen recipe for predicting counts for the longer IRAC bands,
is to take the newly available mid-IR extragalactic counts 
at 6.7 $\umu$m and shift these to the neighbouring IRAC bands. 

We thus fitted an experimental source count slope to the 6.7 $\umu$m 
extragalactic counts found by various
ISOCAM surveys:  ELAIS (Serjeant \ea 2000),
ISO-HDF project (Oliver \ea 1997),
a survey at the Lockman Hole (Taniguchi \ea 1997), and the CFRS (Flores 
et al.\ 1999).  
The faint slope beyond the deepest observations is
adjusted to typical shape of existing (evolutionary) models
(Franceschini et al.\ 1997, Pearson \& Rowan-Robinson 1996, 
Roche \& Eales 1999). 
Preliminary {\em ISO}-counts from the HDF-S field are also
in excellent agreement with this experimental source count model
(Oliver et al. 2001, in preparation).
Though various independent counts are in very good agreement, it should be 
noted that there still are uncertainties in the calibration of the ISOCAM data 
(see eg.\ Serjeant et al.\ 2000, Aussel et al.\ 1999).  
Note also that the ELAIS counts are not corrected for a possible 
Eddington/Malmquist type bias due to a non-Gaussian flux-error distribution 
(see Figure 1.\ in Serjeant et al.\ 2000).

\begin{figure}
\centerline{\psfig{figure=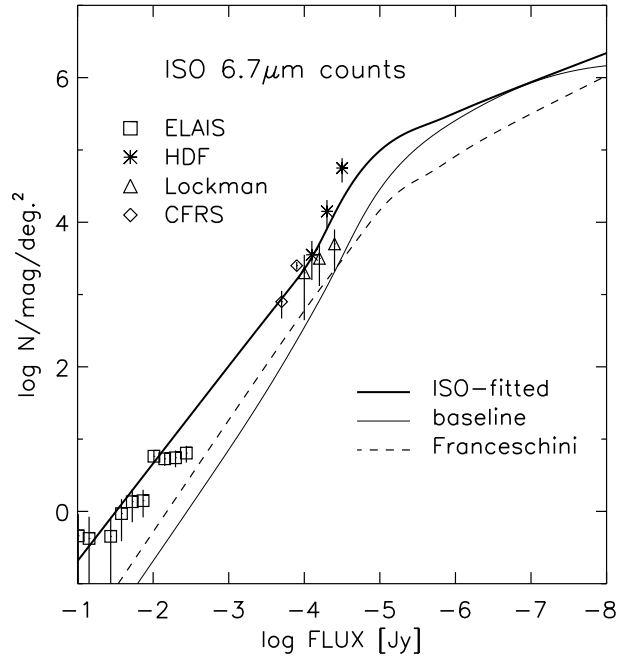,width=9.5cm} }
\caption{Observed differential 6.7 $\umu$m ISOCAM counts:
wide-field ELAIS counts (Serjeant et al. 2000; 
squares), deep HDF-N (Oliver et al.\ 1997; asterisks), Lockman Hole 
(Taniguchi et al.\ 1997; triangles), and CFRS data (Flores et al.\ 1999; 
diamonds).  The thick curve is a fit to these ISOCAM data. 
The thin solid line shows the prediction of
the same baseline model as used in the previous figure.    
The dashed curve is the model from Franceschini et al.\ (1997).}
\label{model67_counts}
\end{figure}

\begin{figure}
\centerline{\psfig{figure=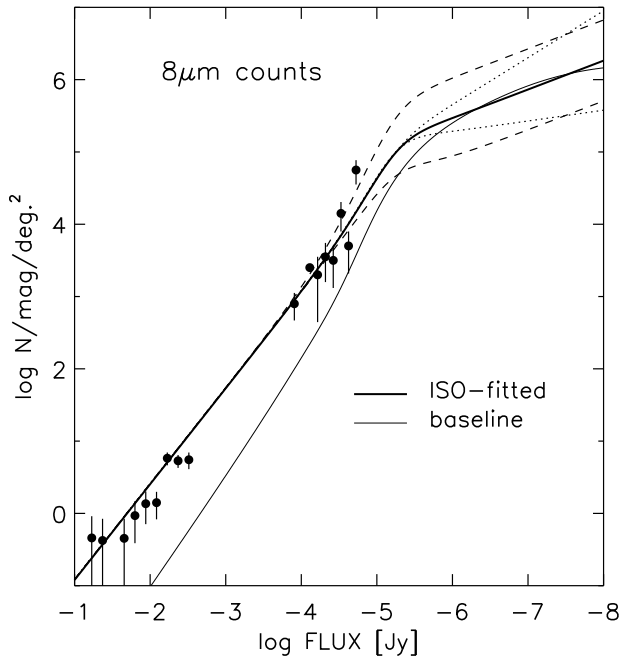,width=9.5cm} }
\caption{Simulated 8 $\umu$m IRAC counts.  The data points are those from 
Fig.~\ref{model67_counts} shifted to the 8 $\umu$m band.   
The thick solid line is translated from the {\em ISO}-data fitted curve of the 
previous figure, and the thin solid line is the baseline model.  
The dashed and dotted lines
show {\em ad hoc} shapes for the faint end of the counts, used to examine 
the effect on confusion limit and the depth reached with the IRAC 
instrument (Section~\ref{disc_irac}).}
 \label{model8_counts}
\end{figure}

Fig.~\ref{model67_counts} shows the recent mid-IR counts 
with our fitted model overplotted as the thick solid curve and the 
equivalent of the 3.6 $\umu$m baseline model as the lower thin solid 
line.  We then translated both the 6.7 $\umu$m counts and the fitted model to 
the 8 $\umu$m band (Fig.~\ref{model8_counts}).

Naturally the translation to the 8 $\umu$m band
is model dependent, but since the 
shift is not large (the ISOCAM filter partially overlaps the IRAC 
8 $\umu$m 
filter), the uncertainty is low, even given the uncertain mid-IR spectral 
features.  We used the baseline model in the transformation. 
We computed numerous model 
counts at 6.7 and 8 $\umu$m, with different LFs, galaxy mixes, cosmologies, 
SEDs (with and without PAH features), and found that all these changes 
in the transformation of counts are  
well within the error bars of the observed counts.    

\subsection{Reliability of models}
\label{rel_models}

As is seen from Fig.~\ref{model67_counts}, 
the observations do not constrain the models fainter than about
20 $\umu$Jy, so the faint counts therefore remain model-based.  However, the 
baseline model provides a realistic slope prediction since it is derived 
from near-IR counts, which extend two decades fainter.  Furthermore, the faint
counts result mostly from high-$z$ objects seen at their redshifted
near-IR rest frames, which are well modelled by the adopted SEDs.

We briefly compare some 6.7 $\umu$m model counts with the models used here.
We have plotted the model of Franceschini et al.\ (1997) in
Fig.~\ref{model67_counts} for reference. 
Models of Franceschini et al.\ (1991) are similar to the baseline curve here.
Pearson \& Rowan-Robinson (1996) models
match our thick solid curve very well at and beyond the break point. 
The evolving model of Roche \& Eales (1999) is very close to our 
experimental ISO-fitted curve.
Overall, the faint mid-IR counts at 6.7 $\umu$m (and especially 
at 15 $\umu$m) are surprisingly high and have a steep slope.  

At brighter flux levels, the 6.7 $\umu$m counts fit 
well the models of Pearson \& Rowan-Robinson (1996), as well as 
those of Roche \& Eales (1999), while they are in clear excess of models 
of \eg Franceschini (1997) and the baseline prediction of our near-IR 
extrapolated model.  

Where does this large difference at the mJy level number
count regime come from?  One explanation is the effect of 
PAH features.  As a test, we calculated the 
baseline model in an alternate way, substituting all the spiral galaxy SEDs 
beyond 1 $\umu$m with mid-IR spectra taken from GRASIL results 
(Silva \ea 1998),
which include the broad mid-IR spectral features.
The effect at 1 mJy is substantial: the 6.7 $\umu$m band there is almost a
factor of 10 increase in counts, raising the baseline model just to 
the level of our experimental counts.  At 8 $\umu$m the increase is a 
factor of 30, which takes the baseline curve slightly higher than the
ISO-fitted experimental curve.
By 10 $\umu$Jy the counts with and without PAH features are within a factor of
two again.  Whether it is appropriate to use such SEDs for the {\em whole} 
population
of disk galaxies is uncertain.  Nevertheless, it is clear 
that the experimental ISO-fitted counts used here are by no means 
unrealistic, and, moreover, the PAH features in mid-IR spectra are a crucial
factor in explaining mid-IR counts.

\section{Deriving confusion}

\subsection{Confusion vs.\ beam size}
\label{conf_vs_beam}

\begin{figure}
\centerline{\psfig{figure=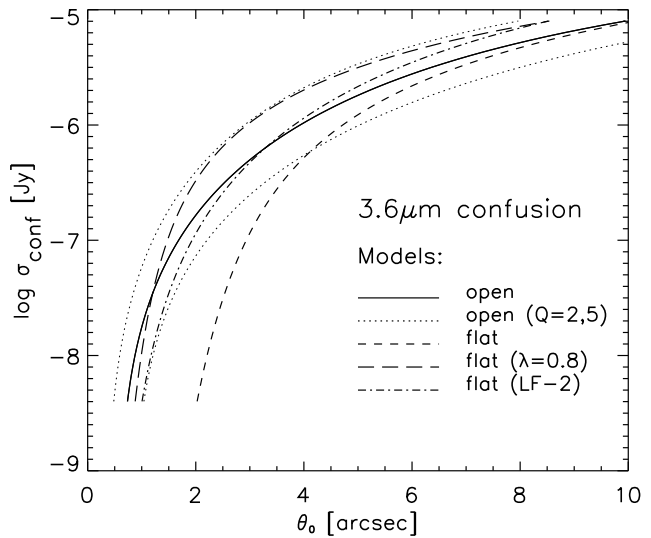,width=9.5cm} }
\caption{Confusion in the 3.6 $\umu$m band
as a function of FWHM of a Gaussian beam.  The solid
line is the calculation based on the baseline model described in the text.
The short dashed and long-dash lines show
results for the same galaxy population and evolution model, but with different
cosmologies (instead of an open model, they use a matter dominated and 
$\Lambda$-dominated flat cosmologies, respectively).  The dash-dot curve
shows confusion using an alternative LF (Szokoly et al.\ 1998).  
The dotted lines
show the effect of the $Q$-parameter in calculation of $\sigma_{\rm conf}$; 
the upper has $Q=5$ and the lower $Q=2$, while $Q=3$ is used for the 
other curves and elsewhere in the paper.}
 \label{theta_sigma3}
\end{figure}

\begin{figure}
\centerline{\psfig{figure=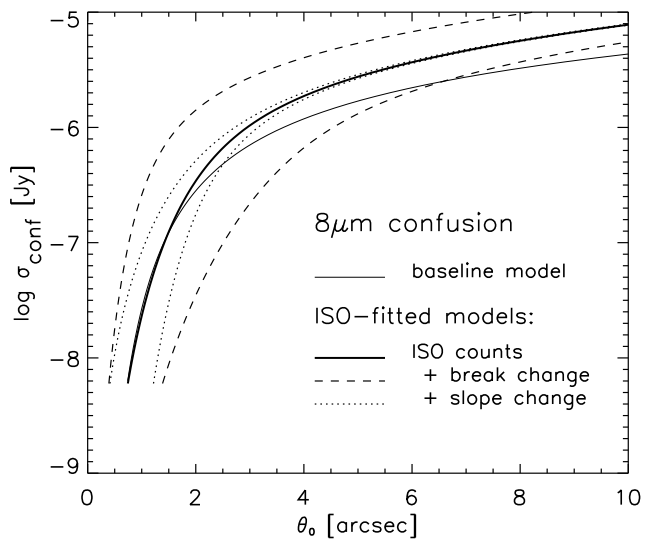,width=9.5cm} }
\caption{Confusion as function of FWHM in the 8 $\umu$m band.  
The source counts which result in these confusion curves are  
Fig.~\ref{model8_counts}. The thick solid 
curve corresponds to counts fitted to the 6.7 $\umu$m ISOCAM counts 
and shifted to
this band.  The dashed and dotted lines show results from {\em ad hoc} models
of faint counts, not constrained by present observational data 
(Section~\ref{disc_irac}).}
\label{theta_sigma8}
\end{figure}

Equation~\ref{eq6-5} is used to compute confusion as a function of 
FWHM of the beam size ($\theta_{0}$).  This is equivalent to 
confusion limit curves in papers of Franceschini \ea (1989, 1991).  Several 
different cases are plotted in Figs.~\ref{theta_sigma3} and~\ref{theta_sigma8},
corresponding to source counts in 
Figs.~\ref{model3_counts} and~\ref{model8_counts}.

Figure~\ref{theta_sigma3} shows effects in the 3.6 
$\umu$m IRAC filter confusion due to typical changes in 
cosmology and of parameters in the 
$\sigma_{\rm conf}$ calculation. 
The confusion noise $\sigma_{\rm conf}$ for the baseline model is
calculated using 
$Q=3$ (see eq.~\ref{eq3} and discussion following it) and plotted as the
solid line.  Confusion from the 
same model is calculated also using $Q=2$ and 5; these are plotted as dotted 
lines, below and above the first curve, respectively.  The first model is 
then modified to flat cosmology ($q_{0}=0.5$, dashed line; it is 
unrealistic, but is included to compare with some earlier determinations) 
and to flat 
cosmology with a cosmological constant ($\lambda_{0}=0.8$; long-dash).  
Finally, the dash-dot curve corresponds to the same alternative local LF
as used in Fig.~\ref{model3_counts}.

Depending on which model one adopts, IRAC/{\em SIRTF}'s 
1$\sigma$ confusion noise 
from the realistic models at 3.6 $\umu$m seems to be 0.2 -- 0.5 $\umu$Jy
for the IRAC resolution of $\approx 2.3$\arcsec.
It is noteworthy, that changing the $Q$ 
value changes the confusion as much as extreme changes in cosmology.  
It is thus important to check which $Q$ value was used when directly 
adopting some confusion calculation for a given mission (eg.\ 
Toffolatti et al.\ 1998 use $Q=5$ when predicting confusion in
far-IR and sub-mm).

Figure~\ref{theta_sigma8} shows various models for the IRAC 
$8\umu$m filter.  
Next we investigate changes in confusion due to different galaxy population 
models.  The solid curve shows the adopted baseline model, equivalent to the 
baseline 3.6 $\umu$m model.  
The corresponding source count can be found 
in Fig.~\ref{model8_counts}.  The baseline model is 
close to the confusion prediction of Franceschini et al.\ (1991), 
which has been used during the {\em SIRTF} planning phases.  
The resulting ISO-fitted 
confusion curve is plotted as the thick solid line.
The expected 1$\sigma$ confusion noise is $\sigma_{\rm conf} \approx 
0.6 \umu$Jy.

It turns out, that the flux levels contributing most to the confusion 
of the longer wavelength IRAC-filter are
just beyond the sensitivity limit of deepest ISOCAM observations.  
Thus the exact amplitude 
and location of the `bump' seen in the mid-IR counts essentially determines 
the expected IRAC confusion limit.  We return to this issue  
in Section~\ref{disc_irac}.

\subsection{Confusion limit from surface densities}
\label{ruleofthumb}

\begin{figure}
\centerline{\psfig{figure=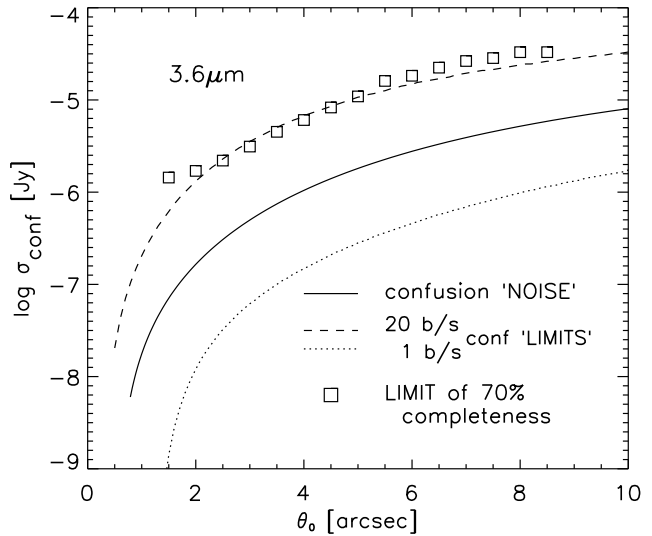,width=9.5cm} }
\caption{Comparison of confusion noise and confusion limit at 3.6 $\umu$m.  
The solid line is the same as 
the baseline confusion noise model in Fig.~\ref{theta_sigma3}.  The 1$\sigma$
confusion limit calculated from one-beam-per-source analysis is shown as
the dotted line, and the corresponding curve for 20 beams/source as the dashed 
line.  Photometric simulations using an actual image (see Section~\ref{compl})
constructed from the same 
baseline model yield 70 per cent completeness limits 
(as a function of the FWHM of
a Gaussian PSF) which are
shown as squares.  It is seen that realistic limits to 
source detection are close to the 20 beams/source limit. }
\label{onebeam3}
\end{figure}

\begin{figure}
\centerline{\psfig{figure=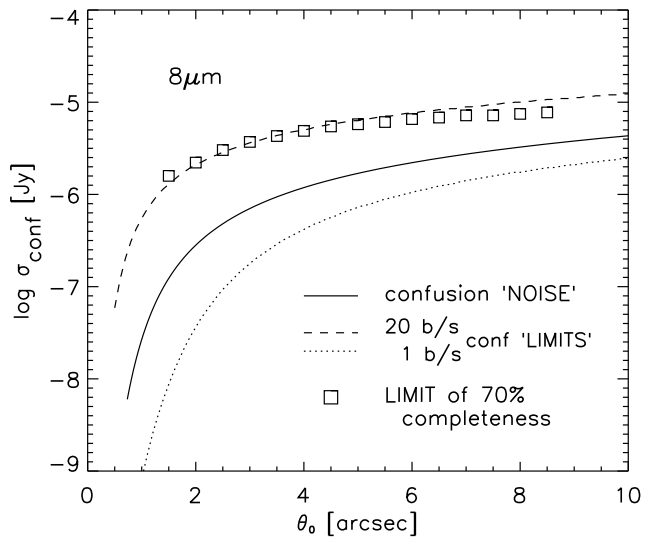,width=9.5cm} }
\caption{Same as previous figure, but for the 8 $\umu$m band.  
The effect of steepness in the 8 $\umu$m 
count slope is seen in the confusion curve: it is flatter over the range of
resolution shown here compared to the 3.6 $\umu$m confusion curve as 
predicted by Eq.~\ref{eq7-5}. }
\label{onebeam8}
\end{figure}

Another often used method 
of defining confusion is to calculate the flux density 
where the surface density of objects per beam becomes unity.
This is done for the same models as above, using
$\omega_{eff} = \pi (\theta_{0}/2)^{2} / \ln 2$ as the approximate
effective beam area (\eg Condon 1974).  
Sets of results for 
both wavebands are shown in Figs.~\ref{onebeam3} and~\ref{onebeam8}.  

The confusion determination from the surface density will be called 
the confusion {\em limit} (dashed and dotted lines).  
In contrast, we will refer to the result of confusion calculation
discussed in the previous section ($\sigma_{\rm conf}$, Eq.~\ref{eq5}),
as confusion {\em noise} (solid lines).  
Both would be called `$1\sigma$' confusion in the literature 
(\eg Blain 1998 et al.), while it is clear that there is a difference 
in the value of inferred confusion. The {\em noise} is a factor of 3 to 5 
higher than the 
{\em limit}, the difference being larger at smaller FWHM's.  
It is unrealistic to adopt
the 1-beam-per-source `rule of the thumb' as an estimate of confusion.
Indeed, often the confusion limit is set to some $N$ beams per source, where
$N$ is in the range 20--50.

Although both of the above confusion limit analyses are often used, they 
are at opposite ends with respect to their approach.  
The analytical calculation presented first 
determines the fluctuations caused by {\em unresolved} 
sources.  The 
second method is based on the realization that the bulk of fluctuations 
resulting from these unresolved sources are caused by sources which have 
a source density of about one per beam.  Using any given (cumulative) 
source count law, one can calculate where that level is, and assign it 
as the confusion limit.  This latter method thus ignores the 
unresolved sources, and instead starts out by calculating numbers 
of bright sources. 
It is thus more related to {\em completeness} analysis in counting of
individual galaxies -- hence the term confusion limit used here.  
Additionally, it is
difficult to associate a meaningful noise figure, one which could be added 
to other noise sources, to this confusion limit; it does not add in 
quadrature like random noise.

\subsection{Sensitivity of confusion to models}
\label{disc_irac}

As seen earlier, the 6.7 $\umu$m model  
is not constrained by observations below $\approx 20$ $\umu$Jy. 
While the faint extrapolation is based on near-IR data and
realistic models,
it is worth-while to investigate how sensitive the confusion noise 
prediction is
to the model-dependent deep counts in the 8 $\umu$m band.
Consider four {\em ad hoc} number count
slopes beyond the {\em ISO}-data regime.  The dashed curves in
Fig.~\ref{model8_counts}  
show models where the amplitude of the bump at the
break point of the counts (at $\sim 10 \umu$Jy) is either higher or lower than 
expected from models fitted to ISOCAM data.  The dotted curves assume 
that the break is determined correctly, but the faint slope is shallower or
steeper than the one adopted previously.

Fig.~\ref{theta_sigma8} shows the calculated $\sigma_{\rm conf}$ for each of
these source counts.  It is obvious that the faint end slope (beyond 
$\approx 6$ $\umu$Jy) does not 
significantly affect the confusion level.  The amplitude of the bump, 
on the other hand, 
is the most important factor contributing to confusion noise.  

The dependence of confusion on the `bump' is expected. 
The 8 $\umu$m sources start to approach beam-per-source 
surface densities with IRAC resolution at about the same region where
the bump, or break, in the counts is.  In addition, 
with steep counts, the sources at the cut-off (or roll-over) level
are the population
contributing most to confusion (eg.\ Toffolatti et al. 1998).  

Basically, depending on whether one normalizes the 
faintest 6.7 $\umu$m counts according to
the HDF-N counts (high; Oliver et al.\ 1997), or to the Lockman hole counts 
(low; Taniguchi et al.\ 1997), one gets almost a factor of 10 difference in 
$\sigma_{\rm conf}$ at IRAC resolution.  The deep ISOCAM observations of 
Altieri et al.\ (1999), through a cluster-lens,  support the
high normalization. 

Nevertheless, 
a word of warning concerning the mid-IR ISOCAM counts is warranted.
Some of the observations have been pushed quite deep, for example, beyond a
20 beams-per-source level.  Our photometric simulations show that
the accuracy of photometry is very low at these levels, and the
trend is systematically towards higher fluxes.  
In addition to unstable photometry, astrometric errors become significant
(Hogg 2000).  Thus, until the counts
are confirmed by better resolution (and hopefully larger sky-coverage), 
it remains a possibility, that the deepest mid-IR counts 
(both at 6.7 and 15 $\umu$m) are significantly over-estimated.  
If that would be the case, the 8 $\umu$m baseline model 
would be more appropriate for estimating confusion levels than the 
ISO-fitted one.

\section{Discussion}
\label{disc}

\subsection{Completeness of counts in IRAC bands}
\label{compl}

Both completeness and reliability must be considered when detecting
faint sources.  Certain detection techniques might give high completeness 
rates while suffering from poor reliability.  On the 
other hand,
requiring high reliability means lower completeness at a given flux level.
We study both aspects, along with the effect of source confusion, 
using simulations of images which are realizations of model
counts introduced earlier.

\begin{figure}
\centerline{\psfig{figure=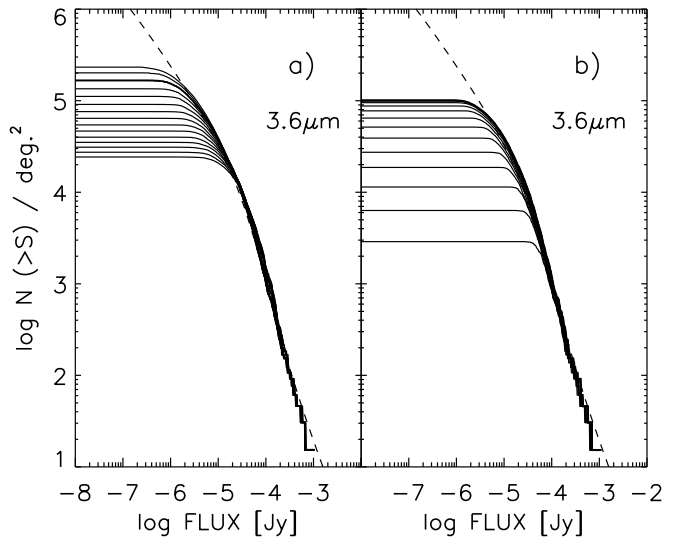,width=9.5cm} }
\caption{Cumulative counts of the 3.6 $\umu$m baseline model (dashed curve)
with `observed' counts from simulations (solid lines).  In {\em a)} 
the curves correspond to different
convolutions: the top-most has a 1.5\arcsec\ FWHM Gaussian PSF, and the rest 
have PSF's larger in steps of 0.5\arcsec\ upto 8.5\arcsec.  The thick
solid line corresponds approximately to the actual IRAC beam size.
In {\em b)} the extracted curves all use a realistic IRAC PSF 
($\theta \approx 2.3$\arcsec) of this band. 
The topmost (thick) line has no noise added. 
The rest of the curves have increasing
amount of random noise added to the image.  See text for details.}
\label{cumcounts3}
\end{figure}

\begin{figure}
\centerline{\psfig{figure=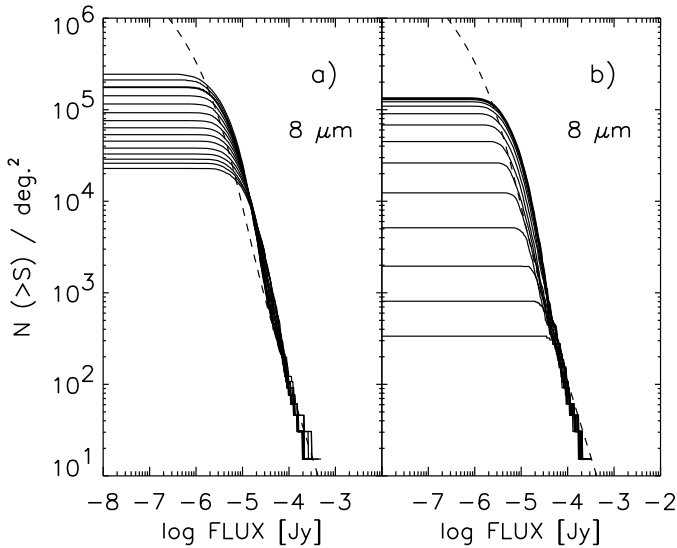,width=9.5cm} }
\caption{Same as previous figure, but for the baseline 8 $\umu$m model:  
in {\em a)} the size of the PSF changes while in {\em b)} the
noise floor is different from image to image. 
If counts are very steep close to the detection limit, 
it becomes increasingly difficult to
measure fluxes without overestimating the true flux.  This shows itself 
as a bump in the extracted counts just before the completeness limit.}
\label{cumcounts8}
\end{figure}

An analogue of the $\sigma_{\rm conf} (\theta_0)$ 
calculation in Section~\ref{conf_vs_beam} 
is to test how the extraction of sources from a 
simulated image changes with the convolving PSF.
We constructed simulated images from the baseline 3.6 and 8 $\umu$m models --
Figs.~\ref{cumcounts3}a and~\ref{cumcounts8}a show the 
3.6 and 8 $\umu$m intrinsic cumulative model counts as dashed curves.  Sources 
were drawn from $N(S)$ and distributed randomly on an 
image with 1.2\arcsec pixels (fractions of pixels were allowed as the
center of a source).  The image size was 768 pixels$^2$, corresponding 
to a 3x3 mosaic of IRAC frames.  
The image was then convolved with different 
Gaussian PSFs and sources were extracted using SExtractor 
(Bertin \& Arnouts 1996) and aperture magnitudes with approximately
the size of the FWHM were calculated.  
The cumulative `observed' counts are overplotted as solid lines.

A comparison to confusion can be made by defining 70 per cent,
for example, as a
satisfactory level of completeness and determining the flux density 
level where this completeness is reached.  These were determined from each
simulation and plotted in 
Figs.~\ref{onebeam3} and~\ref{onebeam8} as squares. 
The required 70 per cent completeness 
correspond to 3--6$\sigma_{\rm conf}$ 
confusion noise levels at FWHMs of 
several arc-seconds and up to $10\sigma_{\rm conf}$ at smaller 
FWHMs.  

The dashed lines in Figs.~\ref{onebeam3} and~\ref{onebeam8} 
show the limits where there are 20 beams per source.  
The completeness limit is close to this level.
Confusion limits are often quoted as being at 30 -- 40 beams/source, which 
is realistic when higher completeness 
is required (or the definition of a beam is smaller, as eg.\ in Hogg 2000).  
Also, the example above showed a very idealized 
case regarding object detection -- 
perfect Gaussian point sources, no additional
noise or source clustering.  We will return to the issue of detection below.

Another way to derive confusion limits from simulations is to test 
how the extraction of sources changes as 
noise is added to an intrinsically noiseless image.  The same 
simulated images as above were used, except that now the real IRAC PSFs 
for the two bands were used (FWHMs for both bands are $\approx 2.3$\arcsec).   
Fig.~\ref{cumcounts3}b shows the same intrinsic cumulative 
counts as earlier and the extracted counts corresponding to different 
amounts of additional noise. Starting from the bottom, 
the noise levels correspond approximately to
the following integration times (IT, 
in seconds): 12, 24,
48, 60, 90, 100, 200, 500, 1200, 3100, 7900, 19800, and 49600 
s.  We have used the low background environment and
the full-array readout mode, which has only
IT's of 12, 30, 100 and 200 s available, each of 
which has its own sensitivity (an IT of 60 seconds results in 
a different noise level depending on whether it is acquired with 5 coads
of 12 s IT's or 2 readouts of 30 s IT's).
The top-most observed curve is the truly 
confusion-limited case, where no noise is added.  It is 
somewhat lower than the case with a Gaussian PSF of approximately the same
size as used in Fig.~\ref{cumcounts3}a (thick line), showing the effect 
of the extended wings in the real PSF.

The corresponding case for the 8 $\umu$m baseline
model is shown in Fig.~\ref{cumcounts8}b. 
The corresponding integration times (in {\em ksec}) 
for the curves starting from bottom are: 0.1, 0.3, 0.7, 1.8, 
4.4, 10.9, 27.1, 67.0, 166, 411, 1020, and 2530 ksecs.  The top-most
curve has already merged with the no-noise curve -- however, it has
an IT of 700 hours, an amount only the largest Legacy Science projects have.

\begin{figure*}
\centerline{\psfig{figure=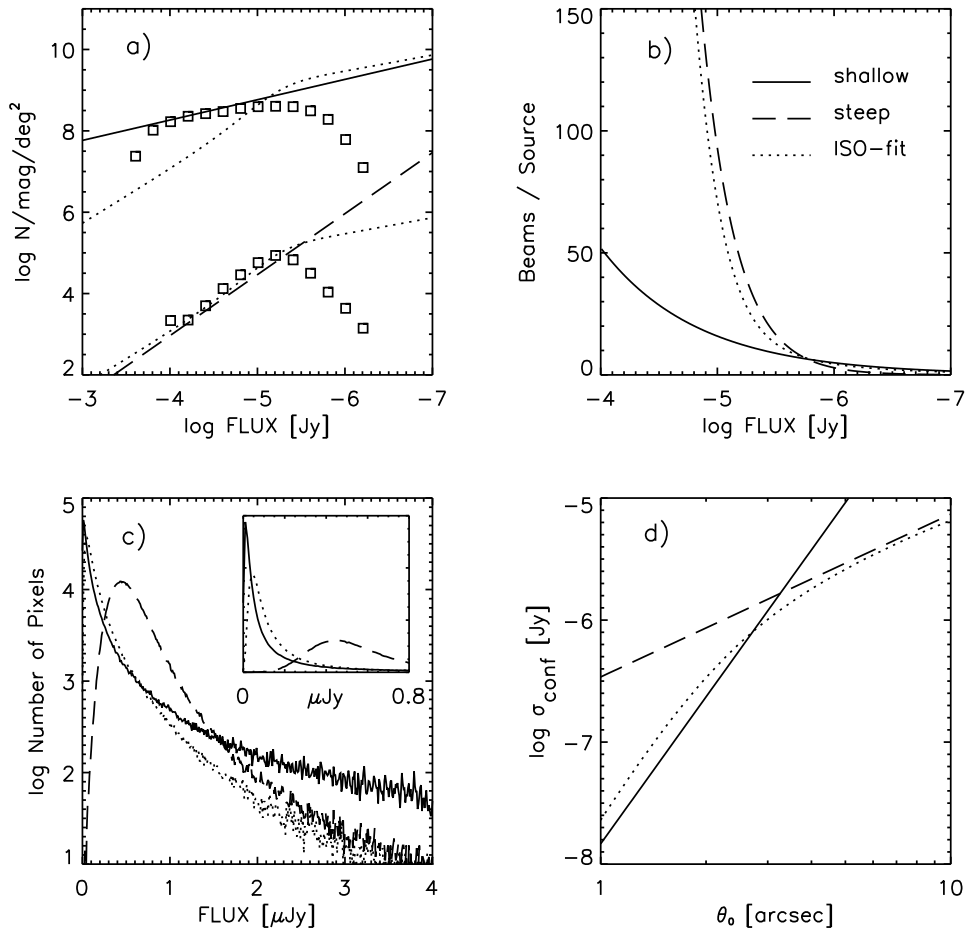,height=13.cm} }
\caption{An illustration of effects of confusion noise and confusion limit 
using power-law source counts. Panel {\em a)} 
shows two source counts, Euclidean
counts as the dashed line, and shallower counts as the solid line (the latter
is multiplied by $10^4$ for clarity).  The dotted line shows the ISO-fitted
8 $\umu$m counts for reference.  Panel {\em b)} plots the 
cumulative surface density as beams/source; 
{\em c)} shows the pixel histrogram (or the total $R(x)$ distribution).
The inset has the same curves (number of pixels as a function of flux) 
but plotted on linear axes to clarify the
low response region; panel {\em d)} 
plots the confusion noise, $\sigma_{\rm conf}$ as a function of beam FWHM.}
\label{example}
\end{figure*}

\subsection{Reliability of detections}

Naturally the extraction algorithm plays a role in construction of
the observed source counts and the exact determination 
of completeness levels.  For example, it is seen that the extracted counts
show an excess compared to input sources in 
Figs.~\ref{cumcounts3} and~\ref{cumcounts8}. 
Examination of input and output sources in our simulations showed that
this bump is mostly caused by
over-estimated fluxes of sources close to the detection limit.   
The over-estimation is due to a `pedestal effect': unresolved 
sources are entering the aperture where the source's magnitude is measured,
and normal sky-subtraction routines are biased because of the non-Gaussian
noise coming from the positive-only signal of faint sources.
We found this to be typical to all photometric systems based on aperture or 
isophotal photometry in very crowded fields.  
PSF-fitting techniques, or extremely small apertures with subsequent 
aperture corrections, give more accurate results 
in point source cases, such as the current test;  however, these methods in
turn have complications with 
real data where other shapes of objects are expected as well.  
The common Malmquist/Eddington-like bias 
is also partially responsible for the bump in the counts. It affects directly
the numbers of sources counted: as there are more objects
per flux (or magnitude) interval below the flux limit than above it, 
photometric errors preferentially scatter more fainter sources to a brighter 
flux bin than {\em vice versa}.

The depth reached can be optimized by tuning appropriate detection 
parameters in the selected extraction method.  
However, a detailed  
study of detection and photometric techniques is out of the scope of this 
paper.  To isolate the completeness vs.\ confusion effects,
we merely took care to be consistent from one test to another using
small aperture magnitudes within SExtractor.  

It is worth pointing out, however, that regardless of the extraction algorithm
used the source count slope and the bright sources have a significant
and large effect on
confusion and completeness.  In the traditional parlance of 
Section~\ref{analytical}, the high response tail of the $R(x)$
distribution strongly affects the detection of sources.  And it is the
slope of the counts which determines $R(x)$.  
To illustrate this point, 
we performed simulations with pure power-law source counts. 

Consider two source counts, showed in Fig.~\ref{example}a. 
The differential counts
have a power-law form $N(S) \propto S^{-\gamma}$, where the shallow curve has 
$\gamma = 1.5$ while the steep one obeys Euclidean counts with 
$\gamma = 2.5$.  Both are adjusted to give counts similar to the ISO-fitted 
$8 \umu$m counts (dotted line) at $\sim 10 \umu$Jy.  These counts are 
urealistic, of course, but do illustrate clearly the dangers of adopting
confusion limits without checking how they are derived.
The squares show the 
observed counts from images made using the two galaxy counts.
The characteristic bump of the observed counts is seen with the steep 
slope.  A steady decrease of completeness is evident with the shallow 
counts: the 70 per cent completeness limit is 
at $10 \umu$Jy while for the Euclidean
counts the limit is closer to $5 \umu$Jy.  The reason for incompleteness
is seen in panel b) where the cumulative surface density of sources is
plotted. There are many bright objects in the shallow count; 50 beams/source
is reached at 100 $\umu$Jy where the completeness starts to drop.  
The 70 per cent completeness is at 17 beams/source.  For the steep 
counts the drop also begins at $\approx 50$ beams/source, but is much faster,
and by $\approx 20$ beams/source completeness correction of a factor of 5
would be needed.
This is because we are approaching confusion {\em noise} instead of 
confusion {\em limit}.

Panel c) 
shows the pixel histograms of the images; these are equivalent to the total 
$R(x)$ distribution (Eq.~\ref{eq2}).  The distribution approaches
a Gaussian as the slope steepens; with shallow slopes the 
histrogram approaches the $N(S)$ distribution (eg.\ Condon 1974).  The width
of the distribution is related to the calculated confusion noise 
$\sigma_{\rm conf}$.  This is shown in panel d) as a function of FWHM.
The $\sigma_{\rm conf} (\theta_0)$ dependence of Eq.~\ref{eq7-5} is also
verified.
It is seen that the $\sigma_{\rm conf} (\theta = 2.3 \arcsec)$ 
from the Euclidean counts is a factor of 2 higher
than from the shallow counts, in direct contrast with the expectation
from source extraction
simulation and surface density analysis above. For the steeper counts one
can assign \eg 5 $\umu$Jy as the confusion limit, which would be equivalent
to $4.5\sigma_{\rm conf}$ or 32 beams/source.  The corresponding
numbers at 70 per cent completeness ($10\umu$Jy) for shallow curve are
$19\sigma_{\rm conf}$ and 17 beams/source. 

It thus turns out that the slope of the 
counts determines which sources dominate the
confusion.  If the source count slope is shallow, it is the bright 
sources which dominate, reducing the completeness of detection 
of an object at a given flux level.  In this case the rule-of-thumb 
taken at $\ga 20$ beams/source (lower for shallow counts and
higher for steeper ones) gives
a good idea of the practical confusion limit.  

If the slope is steep ($\gamma > 2$) the confusion 
noise due to fainter sources starts reducing the completeness faster
than the effect of bright neighbouring sources 
(see also discussion in Helou \& Beichman 1990). In this case the traditional
way of calculating $\sigma_{\rm conf}$ from Equation 5, and using 
$\approx 10\sigma_{\rm conf}$ for the confusion limit is more appropriate.

\subsection{Integration times for IRAC}
\label{inttimes}

\begin{figure}
\centerline{\psfig{figure=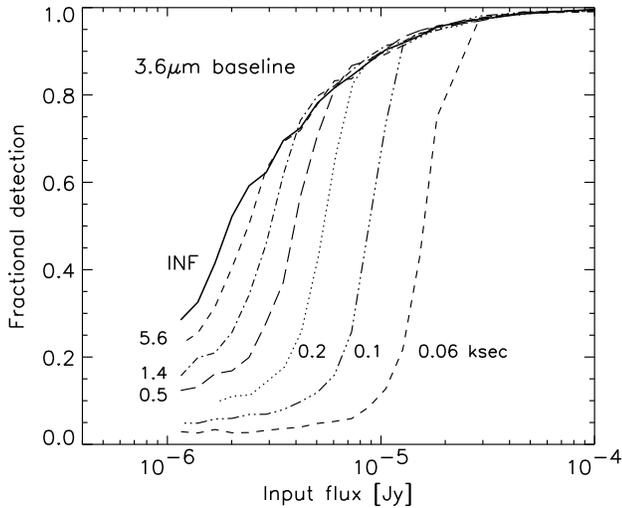,width=9.cm} }
\caption{Fractional completeness level as a function of the flux of the 
input source.  The image in this Monte Carlo simulation is constructed from the
baseline model at 3.6 $\umu$m.  The curves are labelled with the 
integration time
in thousands of seconds. 
The thick solid line is the truly confusion limited case, with no 
noise added to the image, ie.\ infinite integration time. }
\label{completeness3d}
\end{figure}

\begin{figure}
\centerline{\psfig{figure=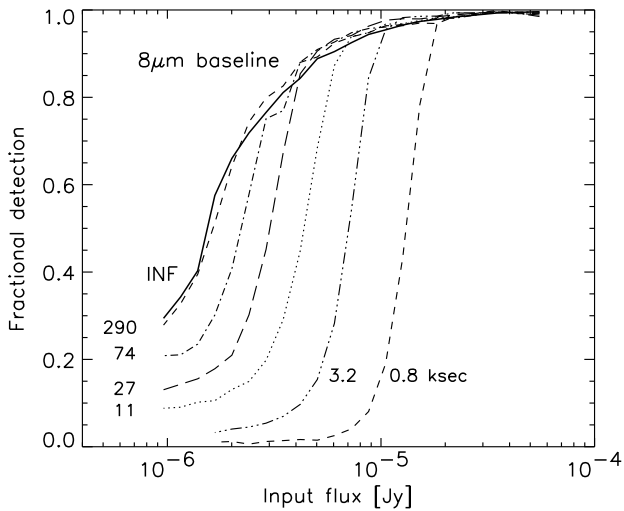,width=9.cm} }
\caption{Same as previous, but for the 8 $\umu$m band baseline model. }
\label{completeness8d}
\end{figure}

\begin{figure}
\centerline{\psfig{figure=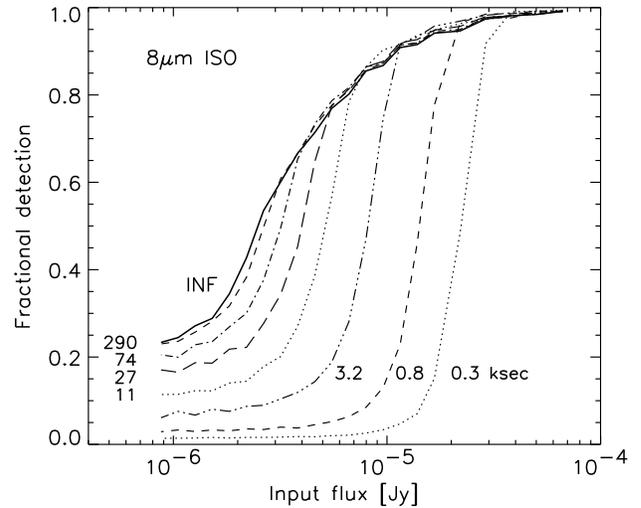,width=9.cm} }
\caption{Same as previous, but using the ISO-fitted model 
to construct the image.  Confusion determined from this plot is 
clearly higher than in the baseline model case.  For example, objects
with intrinsic fluxes of 10 $\umu$Jy can never be detected with 
90 per cent completeness in ISO-fitted model, whereas, if the baseline model
represents the universe, this could be done in about 4 hours. }
\label{completeness8i}
\end{figure}

Figures~\ref{completeness3d}, \ref{completeness8d}, and \ref{completeness8i} 
show the expected completeness levels using different integration times for 
observations with IRAC's 3.6 and 8 $\umu$m filters with the realistic 
PSF's.  Here we used a comprehensive Monte 
Carlo simulation: rather than extracting the source count once from an 
simulated image (as was done for Figs.~\ref{cumcounts3} and~\ref{cumcounts8}),
test sources of 
a given flux were placed one-by-one on the simulated frames and then 
extracted.  In each flux bin for every simulation with different noise floor, 
the test was performed 2000 times.  The integration times 
corresponding  to the different $1\sigma$ noise floors are printed in 
{\em ksecs}.  We show the baseline models and also the
ISO-fitted model for the 8 $\umu$m band.

A `detection' can be defined in various ways. Here it was defined 
as finding a source (the closest one) 
within a FWHM radius and within a $4\sigma$ flux 
interval determined from the observed spread of fluxes at the given flux 
density.  Making the selection in this way we, first of all, avoid a
serious {\em a priori} knowledge bias which would be introduced if 
the exact same input source were searched for from the simulated image.  
Secondly, we chose the mentioned flux cut to be as little as possible
dependent on the used photometric technique.  If, for instance, we had 
required a detection to be within $\pm 30$ per cent of the input flux (this is
used in Hogg 2000) there would have been more non-detections due to the 
typical brightening of sources when approaching the confusion limit.
In this case the resulting no-noise completeness curves in 
Figs.~\ref{completeness3d}--~\ref{completeness8i}
would be lower by a factor of $\sim 1.5$ at a few $\umu$Jy level.

Figures~\ref{completeness3d} to~\ref{completeness8i} can be used to 
estimate the longest useful integration times for the IRAC instrument
in deep extragalactic surveys.  
To be totally confusion limited at 3.6 $\umu$m,
\ie that the incompleteness would be due only to confusion,
ITs of approximately 50 ksec (14 hours) would be needed.  This
is within the scope of a Legacy Science type project, where a `noise
map' of unresolved background sources from a reasonable size of area 
could be obtained for fluctuation studies.  

\begin{table*}
 \centering
 \begin{minipage}{110mm}
 \caption{Confusion limits in $\umu$Jy for two IRAC 
bands using several models. 
Different values of confusion are given according to the method the
limit is derived.  In all the models in this table a value at 
$\approx 20$ beams/source, or $\approx 10\sigma_{\rm conf}$ gives 
a good estimate (`adopted confusion limit') 
of the flux density level which could ultimately be 
reached with accurate enough ($10\sigma$ with ideal photometry) observations.  
However, for a practical limit, the required integration time (IT) 
would need to be considered as well.}
 \begin{tabular}{lccc}
Method & 3.6 $\umu$m baseline & 8 $\umu$m baseline & 8 $\umu$m ISO-fitted  \\
\hline
1-source/beam & 0.02 & 0.07 & 0.06 \\
20-source/beam & 2.0 & 2.3 & 4.2  \\
confusion noise $\sigma_{\rm conf}$ & 0.25 & 0.40 & 0.51 \\
Flux at 70 per cent compl. & 2.2 & 2.3 & 4.1 \\
\hline
Adopted confusion limit & 3 & 3 & 4 \\
Corresponding IT (ksec) & 3  & 150 & 50 
\end{tabular} 
 \end{minipage}
\end{table*}

On the other hand, 
as far as direct source counts are concerned, much shorter ITs are
adequate.  Consider a 500~s exposure. At 6 $\umu$Jy, sources
would still be detected to the confusion limit; that is
sources would be lost at 20 per cent
rate due to confusion only.  Half of all 4 $\umu$Jy objects would be 
undetected;
$\sim50$ per cent of these are lost due to confusion and the other half 
due to sky noise.  This 4 $\umu$Jy level corresponds to approximately
$15\sigma_{\rm conf}$ as calculated from Eq.~\ref{eq6-5}.  Moreover, 
it is easy to
see that the confusion limit is approached; if the IT is increased
by a factor of 10, sources only 1.5 times fainter would be detected
with the same confidence level.  

If only the 3.6 $\umu$m band were considered, we would conclude that it 
does not make much sense to extend the integration time beyond 1 hour.
With this IT 3 $\umu$Jy sources are well detected.
However, one is pushing confusion already since
only $\sim60$ per cent of all sources at that flux level could be extracted 
(it becomes unreliable to interpret and model the count slopes
if completeness corrections exceed a factor of 2).
This limit corresponds to approximately 25 beams per source.

In the longest IRAC waveband, 
with the baseline model, integration times exceeding 10 hours
would still uncover new sources at $3-4 \umu$Jy level 
with only modest completeness corrections of $\sim 1.2$.
However, truly confusion limited images at 8 $\umu$m 
are much harder to reach than at 3.6 $\umu$m -- they would require 
nearly 100 hours of integration time. 

A practical 8 $\umu$m confusion limit is reached at a brighter flux level 
if galaxy counts are better modelled with the ISO-fitted counts. 
Requiring 70 per cent completeness (with $\approx10\sigma$ ideal photometric 
detections), sources at 4 $\umu$Jy would be detected
with the ISO-fitted model, while 2 $\umu$Jy objects could be reached
with the baseline model.  However, in the latter case the IT needed would
exceed the scope of any realistic project.

For the 8 $\umu$m band we conclude that an integration time of about 
15 hours would be both sufficient and useful for performing 
a deep survey (at the same time this would allow truly confusion
limited images from the two shortest wavelength filters).
Sources at 4 $\umu$Jy would be at $8\sigma_{\rm conf}$ or
$11\sigma_{\rm conf}$ depending whether galaxy counts are closer to the 
ISO-fitted or baseline model, respectively.
Respective completeness levels would be at $\approx$ 60 and 80 per cent.
The ISO and baseline models give 20 and 40 beams/source
at 4 $\umu$Jy, respectively.

Table 1 summarizes the integration times along with the expected values
of confusion for the main models used in this work.  It gives the 
confusion limit and noise in the several different methods
presented in this work.
Specifically, it can be seen that the one-beam-per-source limit is nearly
two orders of magnitude too low.  Because the faint mid-IR counts are flatter 
in the $3.6\umu$m band, the overall confusion limit is best described by 
approximately 15 $\sigma_{\rm conf}$.  In the longer IRAC 
waveband the limit
is $\la 10$.  The steepest counts are the baseline $8\umu$m -- the resulting
adopted confusion is approximately 25 beams/source, while for the other two
cases in the table confusion could be described as $\la 20$ beams per source.

There are indications that the IRAC resolution might prove to be better 
than the instrument specifications used here ($2.3\arcsec$).  If this
will be the case, confusion estimates change somewhat,
especially for the two shortest wavelength bands.  
If the 3.6 $\umu$m band would achieve
an in-flight PSF of $\approx 1\arcsec$, the $\sigma_{\rm conf}$ value would
decrease by a factor of 7 (taking into account the 1.2\arcsec pixel size).
A limit of 25 beams/source would be at 0.5 $\umu$Jy.  It would therefore become
difficult to reach confusion limited images in this band.
The 8 $\umu$ band estimates would not change significantly from the
values presented above.  The possibly improving resolution will be 
counteracted by real images, which are not so ideal as the 
point-source simulations performed here but will be spatially resolved 
galaxies.

Finally, note that if 90 per cent completeness were desired, confusion
limits are more appropriately at $\sim 10\umu$Jy, 
or at around 100 beams/source.  If 80 per cent completeness is enough,
IRAC confusion limits are at $\sim 6\umu$Jy.

\subsection{Detection of high-$z$ objects}

The study of high-redshift universe with {\em SIRTF} is highly dependent on 
obtaining accurate enough photometric redshifts of distant galaxies.
This can be performed by tracking the signature of the H$^-$
opacity feature in galaxy spectra as it passes through the IRAC filters 
(Wright, Eisenhardt \& Fazio 1994).
As far as IRAC sensitivities are concerned, 
Simpson \& Eisenhardt (1999) show that a realistic
goal for the deep galaxy surveys
is to determine these redshifts accurately for $L^{\star}$ 
galaxies up to $z\sim4$.  
However, they did not consider confusion in their calculations, 
though they refer to the $1\sigma_{\rm conf} \approx 0.5\umu$Jy expected 
from Franceschini \ea (1991).
The results thus seem optimistic compared to work here.   
Flux densities of 1--2 $\umu$Jy seem impossible to reach.  Certainly 
many individual
sources at this level will be extracted, but they come from deep within 
the confused regime, so 
that obtaining global properties would be highly uncertain
due to large completeness and photometric corrections.

According to our confusion estimates, and using model spectra of an evolved
$L^\star$ galaxy (Fazio \ea 1998), these
galaxies can be detected out to $z \approx 3.5$ using the baseline model 
case.  This is obtained by
setting 3 $\umu$Jy as the lowest flux density level where object 
photometry can be determined with enough accuracy (we assume 
$10\sigma$ in ideal case, which still leaves room for source extraction 
in practice, when objects are not point sources).  In the ISO-fitted
case $z \sim 3$ could still be reached.

The situation changes if the 
highest amplitude ISOCAM galaxy counts (HDF-N counts of 
Oliver et al.\ 1997) prove to be the most accurate counts.   
If the bright bump -model 
(higher dashed line in Figs.~\ref{model8_counts} and~\ref{theta_sigma8})
were used, the 8 $\umu$m band would already become practically 
confusion limited
at 10 $\umu$Jy.  This would limit high-$z$ observation to $z\sim2$.
 
It is just these faint IR-galaxies with unknown 
population characterisctics that {\em SIRTF} projects will be studying.
If the deep surveys detect high counts and surface densities, 
as would be expected from ISOCAM results, 
it would immediately imply strong evolution (\eg Elbaz \ea 1998), 
and possibly hitherto unknown evolutionary stages of infrared-bright objects.  
On the other hand, that would also necessarily mean that the 
IRAC instrument is becoming confusion limited sooner than expected, 
and the highest-$z$ studies of galaxies may suffer.  
Surveys reaching galaxies at $z>3$ are possible only if the 
mid-IR counts are at a moderate level and/or the in-flight IRAC PSF 
performance will be at the most optimistic values in all IRAC filters.

\section{Summary}

We have presented empirically based estimates for confusion due to galaxies
in mid infrared wavelengths. 
For {\em SIRTF}'s IRAC bands the confusion limits are at 
3 -- 4 $\umu$Jy, if completeness corrections by a factor of 1.3 
to intrinsic counts are allowed.  The $1\sigma$ confusion noise ranges
from 0.2 to 0.5 $\umu$Jy moving from the 3.6 $\umu$m waveband to the 8 $\umu$m.
If the highest amplitude {\em ISO}-based 
mid-IR galaxy counts will be confirmed,
the confusion limit is higher by a factor of two.

We have shown the differences between commonly used confusion definitions.
Specifically we separated confusion {\em noise} from confusion {\em limit},
so that the first relates to the fluctuations from unresolved sources and
the latter is defined by source surface density arguments or derived by
completeness analysis of source detections.  The slope of counts determines
which confusion determination is more appropriate to use.  With steep counts
it is the faint source fluctuations which dominate, and when the count slope
is shallow bright sources decrease completeness efficiently and thus determine
the confusion limit.  The combination of 
IRAC resolution and typical mid-IR source counts is an intermediate
case, where the overall 
confusion limit ($\sim 4 \umu$Jy) can be found at 10--20 times the 
confusion noise $\sigma_{\rm conf}$.

We conclude that to make a useful high-redshift ($z\sim 3$) deep survey
with {\em SIRTF}/IRAC, 
integration times of at least 15 hours should be used.  Such a survey would
be confusion limited at 3.6 and 4.5 $\umu$m, if the original IRAC 
design requirement resolution is used.  If sharper images are  
obtained, 3.6 $\umu$m images will become confusion limited at similar 
integration times as the 8 $\umu$m channel, \ie only after 100 hours.

\section*{Acknowledgements}

P.V. wishes to thank the Finnish Academy and the Smithsonian Institution for 
support during this research.

\bsp

\label{lastpage}

\end{document}